%% file: main.tex
\documentclass[runningheads]{llncs}

\usepackage{graphicx}
\usepackage{cite}
\usepackage{amsmath,amssymb,amsfonts}
\usepackage{algorithmic}
\usepackage{graphicx}
\usepackage{textcomp}
\usepackage{tikz}
\usepackage{xcolor}
\usepackage{hyperref}
\usepackage{url}
\usepackage{xspace}
\usepackage{pifont}
\usepackage{listings}
\usepackage{multirow}
\usepackage{bbding}
\usepackage{pifont}
\usepackage{wasysym}
\usepackage{amssymb}
\usepackage{caption}
\usepackage{subcaption}

\input{solidity-latex-highlighting.tex}

\newcommand{\volgapay}{\mbox{\textsc{VolgaPay}}\xspace}

\definecolor{organge}{HTML}{03A60D}

\definecolor{myblue}{HTML}{0026ff}

\begin{document}

\title{System-Wide Security for Offline Payment Terminals}

\author{Nikolay Ivanov \and
Qiben Yan}
\authorrunning{N. Ivanov et al.}
\institute{Michigan State University, East Lansing MI 48824, USA \\
\email{\{ivanovn1,qyan\}@msu.edu}
}

\maketitle

\begin{abstract}
Most self-service payment terminals require network connectivity for processing electronic payments. The necessity to maintain network connectivity increases costs, introduces cybersecurity risks, and significantly limits the number of places where the terminals can be installed. Leading payment service providers have proposed offline payment solutions that rely on algorithmically generated payment tokens. Existing payment token solutions, however, require complex mechanisms for authentication, transaction management, and most importantly, security risk management. In this paper, we present \volgapay, a blockchain-based system that allows merchants to deploy secure offline payment terminal infrastructure that does not require collection and storage of any sensitive data. We design a novel payment protocol which mitigates security threats for all the participants of \volgapay, such that the maximum loss from gaining full access to any component by an adversary incurs only a limited scope of harm. We achieve significant enhancements in security, operation efficiency, and cost reduction via a combination of polynomial multi-hash chain micropayment channels and blockchain grafting for off-chain channel state transition. We implement the \volgapay payment system, and with thorough evaluation and security analysis, we demonstrate that \volgapay is capable of delivering a fast, secure, and cost-efficient solution for offline payment terminals.

\end{abstract}

\input{introduction}

\input{relatedwork}

\input{systemdesign}

\input{implementation}

\input{evaluation}

\input{security}

\input{conclusion}

\input{acknowledgement}

\bibliographystyle{splncs04}

\end{document}

%% file: solidity-latex-highlighting.tex
\usepackage{listings, xcolor}

\definecolor{verylightgray}{rgb}{.97,.97,.97}

\lstdefinelanguage{Solidity}{
	keywords=[1]{anonymous, assembly, assert, balance, break, call, callcode, case, catch, class, constant, continue, constructor, contract, debugger, default, delegatecall, delete, do, else, emit, event, experimental, export, external, false, finally, for, function, gas, if, implements, import, in, indexed, instanceof, interface, internal, is, length, library, log0, log1, log2, log3, log4, memory, modifier, new, payable, pragma, private, protected, public, pure, push, require, return, returns, revert, selfdestruct, send, solidity, storage, struct, suicide, super, switch, then, this, throw, transfer, true, try, typeof, using, value, view, while, with, addmod, ecrecover, keccak256, mulmod, ripemd160, sha256, sha3}, 
	keywordstyle=[1]\color{blue}\bfseries,
	keywords=[2]{address, bool, byte, bytes, bytes1, bytes2, bytes3, bytes4, bytes5, bytes6, bytes7, bytes8, bytes9, bytes10, bytes11, bytes12, bytes13, bytes14, bytes15, bytes16, bytes17, bytes18, bytes19, bytes20, bytes21, bytes22, bytes23, bytes24, bytes25, bytes26, bytes27, bytes28, bytes29, bytes30, bytes31, bytes32, enum, int, int8, int16, int24, int32, int40, int48, int56, int64, int72, int80, int88, int96, int104, int112, int120, int128, int136, int144, int152, int160, int168, int176, int184, int192, int200, int208, int216, int224, int232, int240, int248, int256, mapping, string, uint, uint8, uint16, uint24, uint32, uint40, uint48, uint56, uint64, uint72, uint80, uint88, uint96, uint104, uint112, uint120, uint128, uint136, uint144, uint152, uint160, uint168, uint176, uint184, uint192, uint200, uint208, uint216, uint224, uint232, uint240, uint248, uint256, var, void, ether, finney, szabo, wei, days, hours, minutes, seconds, weeks, years},	
	keywordstyle=[2]\color{teal}\bfseries,
	keywords=[3]{block, blockhash, coinbase, difficulty, gaslimit, number, timestamp, msg, data, gas, sender, sig, value, now, tx, gasprice, origin},	
	keywordstyle=[3]\color{violet}\bfseries,
	identifierstyle=\color{black},
	sensitive=false,
	comment=[l]{//},
	morecomment=[s]{/*}{*/},
	commentstyle=\color{gray}\ttfamily,
	stringstyle=\color{red}\ttfamily,
	morestring=[b]',
	morestring=[b]"
}

%% file: introduction.tex
\section{Introduction}

The popularity of electronic payments has grown significantly over the past decade due to increasing number of online users. Consumers got used to online shopping and payments, however, a number of security incidents have undermined their trust in electronic commerce~\cite{manworren2016you,xu2008security,trautman2016corporate}.

Most electronic payment terminals have to stay online in order to process payments. The online requirement, however, is associated with cybersecurity threats, increased costs, dependency upon third-party infrastructure, and limited locations where the terminals can be installed. Offline payment designs deliver solutions to these problems~\cite{van2009offline,dmitrienko2017secure,jiang2015secure,lind2018teechain,sabba2016token,tebbe2015offline}, which generally require a payment token to be transmitted between the payer and payee through Near-Field Communication (NFC), Bluetooth, electromagnetic field~\cite{mendoza2016samsung}, QR code~\cite{nseir2013secure} or even audio signal~\cite{alipayqr}. 

\begin{figure}
\captionsetup[subfigure]{justification=centering}
\centering
\begin{subfigure}{.47\textwidth}
  \centering
  \includegraphics[height=1.5in]{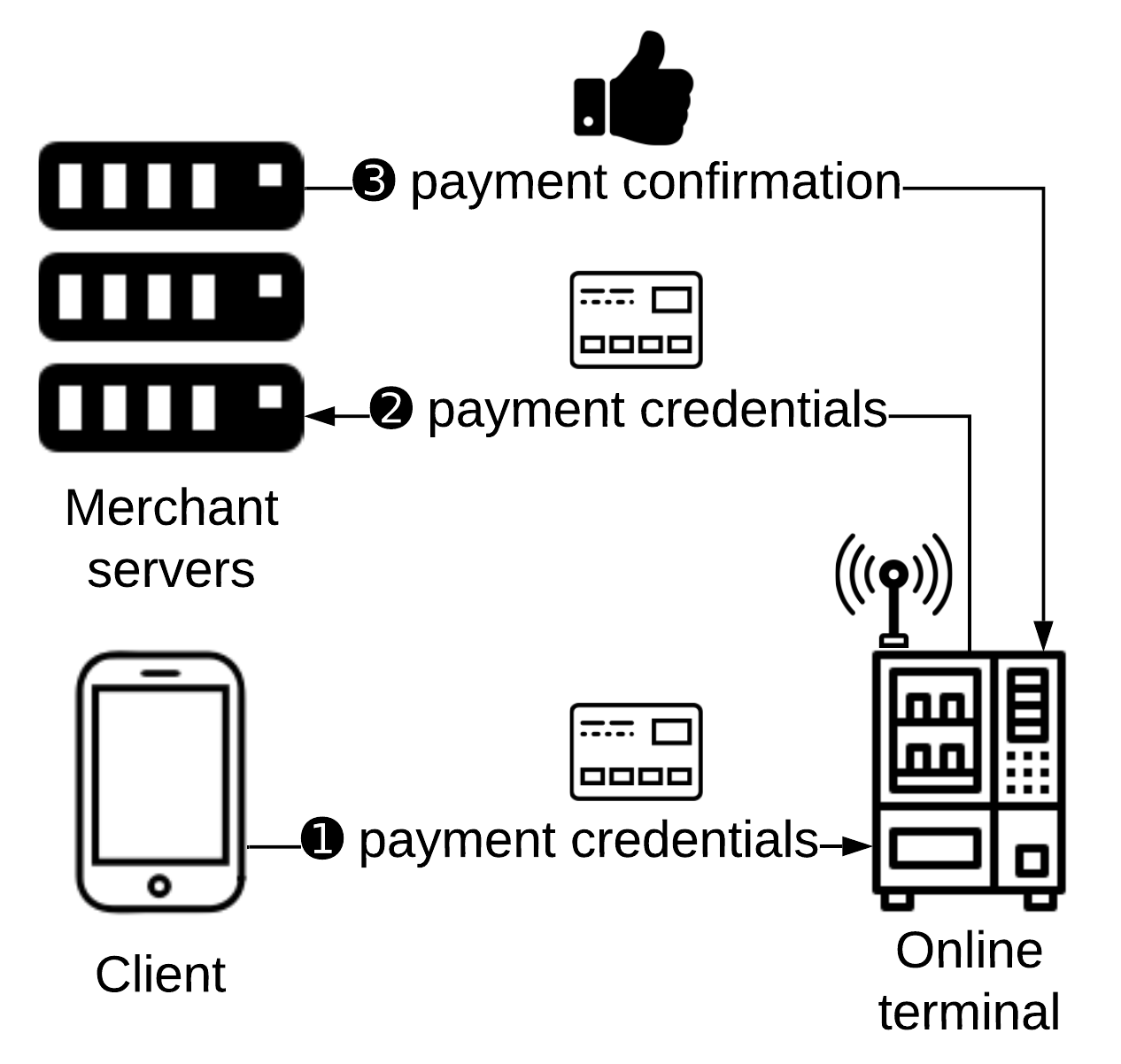}
  \caption{Online terminal}
  \label{fig:offlineonline1}
\end{subfigure}%
\begin{subfigure}{.47\textwidth}
  \centering
  \includegraphics[height=1.5in]{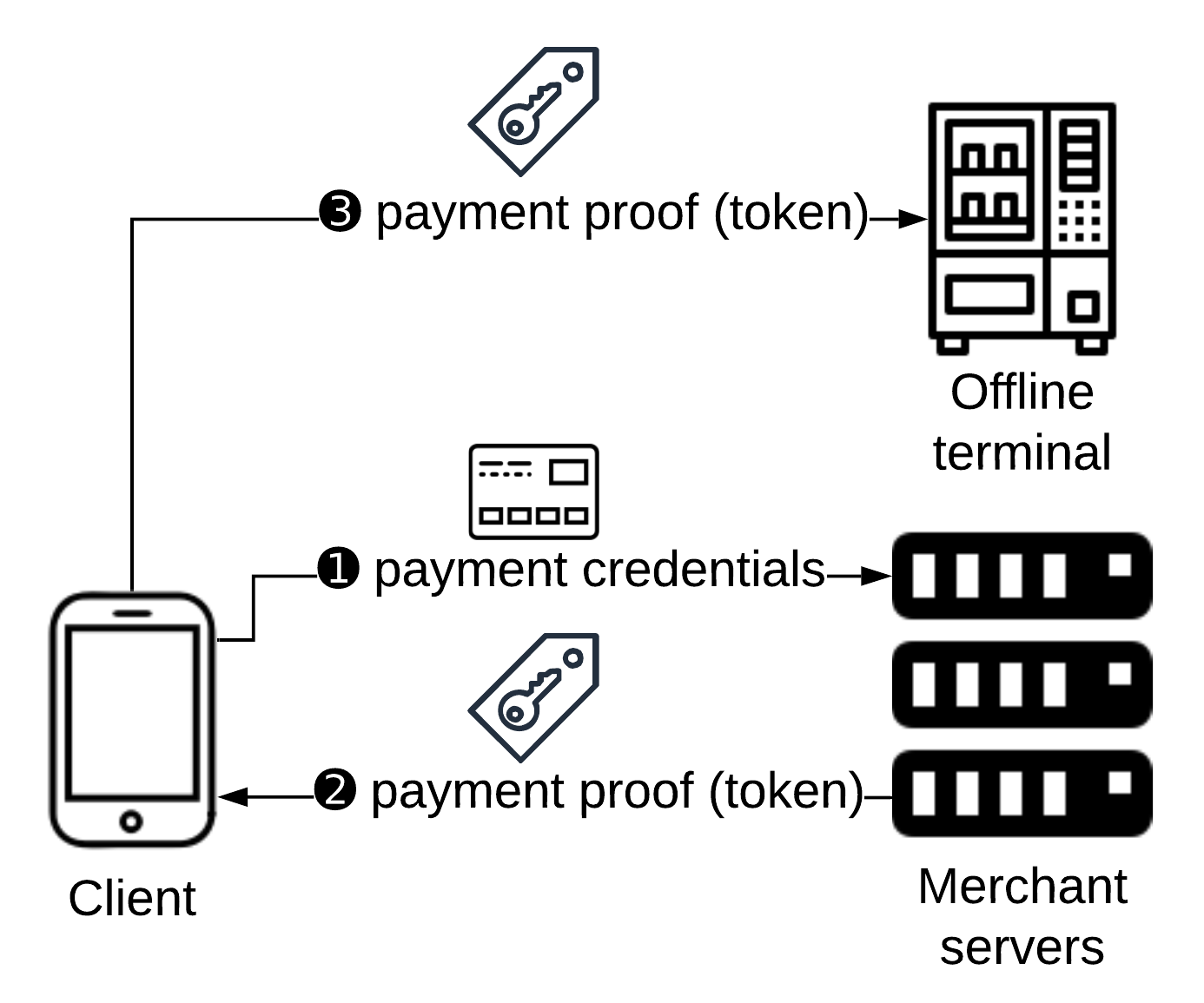}
  \caption{Offline terminal}
  \label{fig:offlineonline2}
\end{subfigure}
\caption{Payment transaction workflows with online and offline terminals.}
\label{fig:offlineonline}
\end{figure}

Figure~\ref{fig:offlineonline} illustrates the payment transaction workflows in payment systems with online and offline terminals. An online terminal gathers client's payment credentials, and uses them for authentication and payment validation at the merchants' server. By comparison, a payment transaction with an offline terminal allows the client to request the payment token from merchant servers directly and use it for payment verification on the offline terminal.

However, one major challenge of designing offline payment infrastructure is to securely manage user balance, including over-spending and double-spending prevention. In traditional payment channels, an offline terminal cannot be timely updated on changes of a user balance, and thus it is subject to token lifting and spending attacks~\cite{bai2017picking}. One way to address this challenge is to use cryptographic payment tokens (i.e., cryptographic proofs of payment) that a user delivers to the payment terminal. Yet, the token-based payment systems face four major concerns: security, privacy, trust, and cost-inefficiency.

First, in the classic payment token designs, the user balances and transactions are stored in a centralized database, which is a single point of failure. The resilience of a payment token infrastructure can be enhanced at the expense of storing multiple copies of encryption and signature keys on different nodes. Second, the sensitive client information, such as credit card number, often has to be stored in the merchant infrastructure, which is notorious for being a target of massive cyberattacks~\cite{manworren2016you}. Third, the classic payment token designs further assume the trustworthiness of the merchants, while the necessity to protect clients from potentially malicious or fake merchants is largely ignored. Fourth, it requires a significant effort and investment to ensure the security and resilience of a classic solution. As a result, the cost and complexity of the payment token infrastructure may outweigh the benefits of using offline terminals.

In this paper, we design a novel payment system, \volgapay, for token-based offline terminals. The design of our system relies on the following two insights:

\begin{enumerate}

\item We use blockchain and cryptocurrency, which provide an effective solution for eliminating otherwise unnecessary reliance of client upon merchant's trustworthiness. We supplement the blockchain technology with a multi-hash chain-based micropayment channel and polynomial price representation in order to achieve a balance of security and efficiency for payment transactions between merchant and client.

\item To achieve high efficiency, enhanced security, and operation cost reduction, we propose the \emph{blockchain grafting} technique to conjoin a fast, zero-fee auxiliary smart contract in a private blockchain with a decentralized smart contract on a public blockchain. 
\end{enumerate}

\volgapay operates offline points of sale (OPOS) and uses multi-hash chain micropayment channels and \emph{blockchain grafting} to enhance the security and efficiency of offline payment transactions. 
Each merchant-client association is represented by a novel combination of smart contracts: one smart contract deployed on the public blockchain (public smart contract), and another one deployed on the private blockchain (private smart contract). The public smart contract provides protection of client's deposit and merchant's payoff without any pre-existing trust, while the private smart contract provides secure interface and accounting to the off-chain micropayment infrastructure of the merchant with minimum delays and zero fees. 

We implement a \volgapay prototype system using Raspberry Pi-based terminals, Android clients, Ethereum-based public and private blockchains, and a cohort of independent token signer servers scattered across the globe. \textbf{A video demonstration of \volgapay operations is available at \url{https://youtu.be/rjIhDD2yi5I}}. 

In summary, our contributions are as follows:

\begin{itemize}
    \item We propose a blockchain-based transaction network, \volgapay, to address the elevated security risks, inefficiency, and high cost of existing payment systems with offline terminals;
    \item We implement \volgapay, which we thoroughly evaluate to determine its low cost, high speed, and impressive scalability: the token request delay normally does not exceed 12 seconds, the one-time blockchain fee of establishing a smart contract is less than \$1, the communication overhead is light enough to serve 3G-connected clients, and the number of simultaneous token requests can reach 10,000.
    \item We analyze the security of \volgapay, and prove that: if any component of the system is compromised, the harm will be localized, and the system at large remains secure.
\end{itemize}

%% file: relatedwork.tex
\section{Related Work}

\noindent \textbf{Off-chain payment channels.} 
The Bitcoin Lightning Network protocol~\cite{poon2016bitcoin} uses Bitcoin smart contract functionality for securing off-chain micropayment channels. To perform off-chain operation, Bitcoin Lightning uses two-way multi-signature channels based on digital signature schemes, supported by a network of multiple dedicated servers. Currently, there are a few test implementations of the Lightning network, including Lightning Labs~\cite{lightninglabs}, Casa Node~\cite{casanode}, ACINQ~\cite{acinq}, Blockstream~\cite{blockstream}, and MIT lit~\cite{mitlit}. Other cryptocurrency micropayment projects include Perun~\cite{dziembowski2019perun}, Revive~\cite{khalil2017revive}, Sprites~\cite{miller2017sprites}, Bolt~\cite{green2017bolt}, SpeedyMurmurs~\cite{roos2017settling}, Stellar~\cite{stellar}, and a duplex micropayment channel~\cite{decker2015fast}. Blockumulus~\cite{ivanov2021blockumulus} uses a cloud-based smart contract, called \emph{FastMoney}, for off-chain payments. The functionality and security of all the solutions above largely depend on the setup and performance characteristics of centralized or crowd-sourced servers that serve the off-chain network. To avoid security risks and synchronization complexity associated with centralized micropayment management, \volgapay uses \emph{blockchain grafting} for transaction management. Moreover, the aforementioned solutions are not designed to support payments with offline terminals.

\noindent \textbf{Offline electronic payments.} 
There are a few solutions proposed for offline electronic payments. The most popular schemes involve delivering a cryptographically verifiable token from client to terminal using a local offline channel, e.g., electromagnetic or radio transmission~\cite{jiang2015secure}. Other solutions propose the use of trusted hardware \cite{dziembowski2018general, lind2018teechain}, which relies on the assumption of hardware integrity. While trusted hardware is useful in a deeply regulated and controlled environment, such as Apple Pay, it could not support anonymous payments in a low-trust environment, e.g., a self-service station.

%% file: systemdesign.tex
\section{System Design}\label{section:systemdesign}
\begin{figure}
    \centering
    \includegraphics[width=4in]{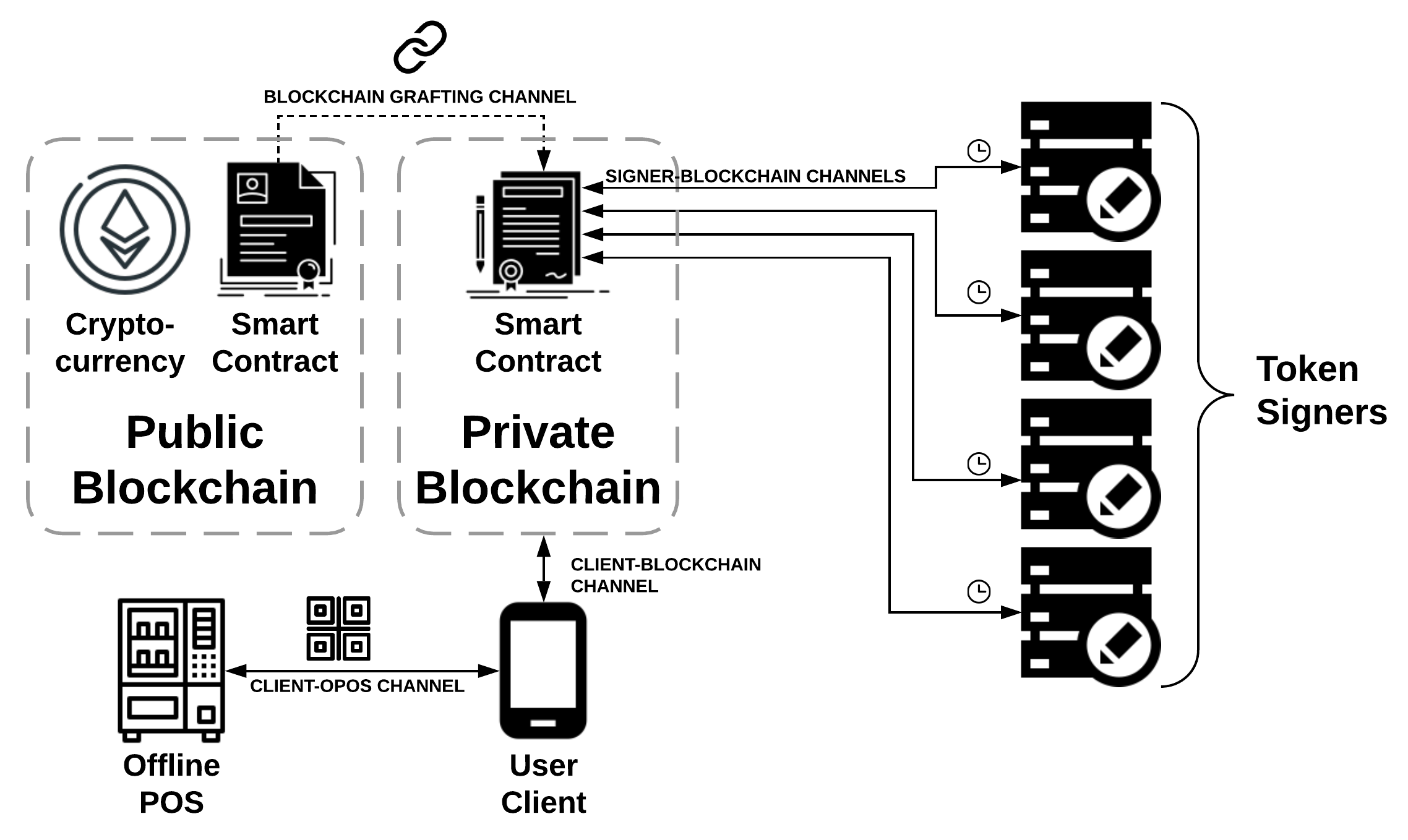}
    \caption{\volgapay system architecture. The \textbf{\clock}~label symbolizes an event-listening channel. \volgapay includes seven types of participants and four types of channels. OPOS first delivers the information needed for payment and token request to the client through the client-OPOS channel. The client then provides payment and requests payment token from the private blockchain through the client-blockchain channel. Signers subscribe to the token request event via the signer-blockchain channels. Private smart contract extends the public smart contract through a virtual blockchain-grafting channel. Finally, the client delivers a payment token to OPOS using the client-OPOS channel.}
    \label{fig:design}
\end{figure}

We elaborate on the design of \volgapay by describing its fundamental definitions and principles, as well as three classes of elements constituting its topology: \textit{participants}, \textit{communication channels}, and \textit{protocols}. Figure~\ref{fig:design} shows the overall system architecture, in which each transaction between client and merchant is supported by a public smart contract supplemented by a linked (grafted) private smart contract. 
The public smart contract protects client's deposit and refund, while guaranteeing payoff for the merchant. The private smart contract provides secure interfacing endpoint for the client and secure accounting and ledger-keeping for the merchants' micropayment channels. The client pays the merchant with a set of hashes, and receives a payment token (i.e., verifiable payment receipt), which is a tuple of signatures from several independent signers that respond to the token request blockchain event. Table~\ref{table:symbols} summarizes all notations used in the design of \volgapay.

\begin{table}
\caption{Notation table for \volgapay.}
\label{table:symbols}
\begin{center}
\begin{tabular}{|c|c|}
\hline
\textbf{Symbol} & \textbf{Meaning} \\
\hline
$\eta$ & Transaction ID randomly generated by OPOS \\
\hline

$\delta$ & Item's listed price or single payment amount \\
\hline

$D$ & Human-readable item description \\
\hline

$N$ & Total number of token signers \\
\hline

$M$ & Number of signatures in a token \\
\hline

$\Gamma_{i}(m)$ & Digital signature of message $m$ signed by token signer $i$ \\
\hline

$K(m)$ & Digital signature of message $m$ signed by an OPOS \\
\hline

$E(p)$ & Private blockchain event with parameter $p$ \\
\hline

$\tau$ & Price base---the sum of previous payments. \\
\hline

$\phi$ & 
$\tau$-adjusted price (\emph{cumulative payment}): 
$\phi = \tau + \delta$\\
\hline

$\Theta$ & Hash chain polynomial base (max. number of hashes in each hash chain) \\
\hline

$\Upsilon$ & Ordered sequence of payment hash chains $\Upsilon = (\Upsilon_{1}, \Upsilon_{2},..., \Upsilon_{\xi})$ \\
\hline

$\xi$ & Number of hash chains in $\Upsilon$ \\
\hline

$\Upsilon_{i}$ & $i$-th hash chain in $\Upsilon$ \\
\hline

$R_{i}$ & The number of hashes released by the client to the merchant from $\Upsilon_{i}$ \\
\hline

$g$ & Price granularity measured in number of atomic units \\
\hline
\end{tabular}
\end{center}
\end{table}

\subsubsection{Polynomial Multi-Hash Chain Micropayment}\label{subs:polymult}

Traditional blockchain-based micropayment channels rely on digital signature schemes, which require the protection of private keys and prevention of double-spending complicated by transaction reversals, whereas hash chain-based micropayment channels eliminate the use and storage of private keys and avoid transaction reversals. In this paper, we use hash chain-based micropayment channel for \volgapay's off-chain transactions. The client generates the hash chains and keeps their seeds in secret. However, the use of hash chain-based micropayment channels involves significant computational overhead. Specifically, long hash chains, which have been successfully used in non-blockchain micropayments~\cite{anderson2017system,pedersen1996electronic}, incur large delays, intolerable fees, and execution timeouts when used in non-view blockchain smart contract calls\footnote{In smart contracts, a ``view" function is a read-only function of smart contract that does not modify the state of blockchain. The ``view" functions can be executed by the blockchain API layer rather than miners. Thus, the execution is generally fast.}. Our pre-evaluation of hash chain generation performance on Ethereum Mainnet has demonstrated significant performance degradation and fee increase for non-view verification of Keccak256 and RIPEMD-160 hash chains whose length exceeds 100 hashes\footnote{On Mainnet, the fee for 100-hash verification exceeds 80\cent~with recommended gas price. The attempt to verify 270 hashes fails with ``out of gas" error.}.

Traditional currencies, such as U.S. dollar (USD), have a coarse granularity (i.e., \emph{cent}) compared to the ultra-fine granularity of most cryptocurrencies, e.g., in Ethereum, 1 \emph{wei} = $10^{-18}$ \emph{Ether}. Moreover, if we assume the granularity of 1 \emph{cent}, the traditional hash chain-based micropayment channel, in which one hash represents one atomic payment unit, would require to produce 10,000 hashes for the verification of a 1,000-\emph{dollar} payment, which is practically infeasible on blockchain due to its significant computational overhead.

To address this problem, we propose a \emph{polynomial multi-hash chain micropayment scheme} which utilizes several hash chains to process arbitrary micropayments. Instead of using a very long single hash chain, in which each hash represents a minimal payment unit, such as Ethereum's \emph{wei}, we use a $\xi$-tuple of short hash chains, each responsible for a single digit with radix $\Theta$. For example, to account for USD prices between 1\cent~and \$10,000 (or $10^{6}$\cent), we need 7 decimal digits, i.e., $\xi = 7,~\Theta = 10$, with 7 hash chains and 10 hashes in each chain. Each digit from 0 to 9 can be represented by one of the 10 hashes, totalling 70 hashes for any price in the given range. In comparison, $10^{6}$ hashes are required in the classic single hash chain representation. Therefore, any single payment amount $\delta$ can be represented as the following polynomial:
\begin{equation}
   \delta = \sum_{i = 1}^{\xi}{R_{i} \times \Theta^{i-1}}, 
   \label{micro1}
\end{equation}
where $R_i$ is the number of hashes released from hash chain $\Upsilon_{i}$. For instance, suppose the client pays \$25.31 (or 2,531\cent), the micropayment will be executed by releasing  1 hash from $\Upsilon_{1}$ ($R_{1} = 1$), 3 hashes from $\Upsilon_{2}$ ($R_{2} = 3$), 5 hashes from $\Upsilon_{3}$ ($R_{3} = 5$), and 2 hashes from $\Upsilon_{4}$ ($R_{4} = 2$), while $R_{5} = R_{6} = R_{7} = 0$.

If \emph{wei} is used as the payment unit, we need 19 hash chains to represent prices up to 9 \emph{Ether}, as 1~\emph{Ether} = $10^{18}$~\emph{wei}. The number of hash chains can be reduced by increasing the granularity $g$ and capping the maximum cumulative off-chain balance. To process multiple payments, we submit the $\tau$-adjusted price (or cumulative payment) $\phi$ instead of the single payment $\delta$, and we use the price base $\tau$ to track the previous payments corresponding to already released hashes. Thus, we adapt Eq.~(\ref{micro1}) to include $g$ and $\tau$. For each payment, the client determines a sequence of \emph{hash chain depths} $(R_{1},\ldots,R_{\xi})$, to express $\phi$ as:
\begin{equation}
\phi=\tau + \delta = g \times \sum_{i = 1}^{\xi}{R_{i} \times \Theta^{i-1}}.
\label{micro2}
\end{equation}
Increasing granularity $g$ reduces the number of hash chains in the set, and thus reduces the public blockchain fees. In our prototype, we use $g = 10^{13}$, which  can represent a range of prices between 0.00001 and 99.99999 \emph{Ether} using just 7 hash chains with 10 hashes for each chain. 

\noindent \textbf{Example:} 
Although \volgapay uses \emph{Ether}, we focus on USD in our examples for ease of illustration. 
Suppose we want to represent USD prices up to \$999,999, with granularity $g = 10^{2}$, i.e., we round the cost to the nearest dollar. Thus, the granularity-adjusted number of hash chains $\xi$ is 6 rather than 8.

Suppose the client makes three consecutive payments: \$1,720, \$56, and \$56. Assume there are no previous payments in the channel (i.e., $\tau = 0$), then for the first payment, $(R_{1}, R_{2},...,R_{7})$ is $(0, 2, 7, 1, 0, 0, 0)$. For the second payment ($\delta = 56$, $\tau = 1,720$, $\phi = 1,776$), the 7-tuple hash chain depth sequence will become: $(6, 7, 7, 1, 0, 0, 0)$. Comparing the two sequences of sets, we can see that in the second payment, the client reveals 6 hashes from $\Upsilon_{1}$ and 5 more hashes from $\Upsilon_{2}$. For the third payment ($\delta = 56, \tau = 1,776$, $\phi = 1,832$), the 7-tuple hash chain depth sequence is $(2, 3, 8, 1, 0, 0, 0)$. Compared to the previous payment, the client releases 1 more hash from $\Upsilon_{3}$. It is worth noting that after three payments, the number of hashes known by the merchant for each hash chain is $(6, 7, 8, 1, 0, 0, 0)$, worth \$1,876. A malicious merchant can try to get payoff of \$1,876 (rather than \$1,832), which will be rejected by payoff verification as illustrated in Section~\ref{subsection:protocol}. 

\subsection{\volgapay Participants}

\volgapay system includes seven classes of participants described below.

\noindent\textbf{Public Blockchain}
Public blockchain is used for storage, access, execution, and integrity assurance of the public smart contract. \volgapay design requires the public blockchain to be universally available, trusted by all parties, capable of executing Turing-complete smart contracts with integrated cryptocurrency.

\noindent\textbf{Public Smart Contract}
Public smart contract is used in \volgapay to establish an unambiguous and non-repudiable agreement between the merchant and the client. Unlike the deposits in classic electronic payment systems, smart contract allows the client to retain full control over deposited funds irrespective of the merchant's trustworthiness. Each merchant-client pair requires a deployment of at least one separate public smart contract.

Table~\ref{table:pubscroutines} describes a minimal set of routines that the public smart contract should have. The \emph{set heads} routine, which saves all hash chain heads at once, can only be called by the merchant. The \emph{payoff} routine exchanges the hashes released by the client into the cryptocurrency funds deposited by the client earlier. The \emph{refund} routine allows the client to request unclaimed funds after a certain time period. The \emph{deposit} routine allows the client  to fund the smart contract.

\begin{table}
\begin{center}
\caption{Minimal set of routines in \volgapay public smart contract.}
\label{table:pubscroutines}

\begin{tabular}{|c|c|c|}
\hline
\textbf{Routine}&\textbf{Description}&\textbf{Access} \\
\hline
Set heads & Store hash chain heads provided by client& merchant \\
\hline

Payoff & Transfer funds to merchant based on cryptographic proof & merchant \\ 
\hline

Refund & Request refund after a payoff or timeout & client \\
\hline

Deposit & Deposit funds on the smart contract & client \\
\hline

\end{tabular}
\end{center}
\end{table}

\noindent\textbf{Private Blockchain}
Unlike its public counterpart, the private blockchain provides security only for the merchant infrastructure, and is not intended to be trusted by the client. Managed by the merchant and executed under a proof-of-authority (PoA) consensus, the private blockchain achieves high levels of security and redundancy for the merchant payment infrastructure at very little cost. Unlike proof-of-work (PoW) consensus, in which the voting power is restricted by computational limits, the PoA consensus limits the voting power by hard-coding the public keys of the sealers in the genesis block. PoA substitutes mining with sealing, and therefore achieves a better performance and full control over consensus. As an option, several merchants can share one private blockchain based on mutual trust. Alternatively, a merchant can use a third-party PoA blockchain as a service delivered by a trusted provider\footnote{Our \volgapay prototype uses Kovan Testnet as a service.
}.

\noindent\textbf{Private Smart Contract}
For each public smart contract, the merchant deploys a mirroring smart contract on its private blockchain. The public smart contract contains an address of the mirroring smart contract in the private blockchain. We call this technique \textit{blockchain grafting}\footnote{Grafting is an agricultural technique, in which a part of one plant is conjoined with another plant in order to combine the benefits of both. Similarly, we join public and private blockchains to utilize the properties and strengths of the two.}. 

 \textit{Blockchain grafting} supplements the public smart contract with the private one, the benefits of which include: 
first, private smart contract can perform secure execution without paying any blockchain fees;
second, private smart contract executes and confirms transactions much faster than its public counterpart; third, private smart contract provides locally-trusted execution and storage within multiple nodes of the merchant infrastructure, such that if one or even several nodes are compromised, the integrity of the ledger remains intact; fourth, private smart contract can achieve secure and flexible redundancy management at low cost: nodes can be added, removed, replaced or updated without service disruption. 

Table~\ref{table:privscroutines} describes the minimal set of routines provided by the private smart contract. The \emph{set heads} routine mirrors its public blockchain counterpart: it sets the hash chain heads to exactly the same values as they are in the public smart contract. The \emph{token request} routine verifies the legitimacy of the OPOS and client, checks the hashes and payment amount provided by the client, processes the payment, updates the list of revealed hashes, and finally triggers a blockchain event that activates the token signers. The \emph{one-time deposit} routine funds a smart contract only once (using a control variable); the additional deposits are prohibited to avoid tampering with the private smart contract balance. The \emph{set $\Gamma_i$} routines are used by the signers to store their signatures in the blockchain.

\begin{table}
\begin{center}
\caption{Minimal set of routines for \volgapay private smart contract.}
\label{table:privscroutines}

\begin{tabular}{|c|c|c|}
\hline
\textbf{Routine}&\textbf{Description}&\textbf{Access} \\
\hline
Set heads & Store hash chain heads provided by client & merchant \\
\hline

Token request& Process micropayment transaction and generate signatures & client \\ 
\hline

One-time deposit & Deposit pseudo-cryptocurrency equal to off-chain balance & merchant \\
\hline

Set $\Gamma_{i}$& Save signature $\Gamma_{i}(\eta, \delta)$ in the smart contract& signer $i$ \\
\hline

\end{tabular}
\end{center}
\end{table}

\noindent\textbf{Token Signers}
To achieve security redundancy, we define the \textit{payment token} as a set of $M$ public-key signatures $\{\Gamma_{1}(\eta, \delta),...,\Gamma_{M}(\eta, \delta)\}$ produced by $M$ out of $N$ signers. The set of signers' public keys is pre-determined by the merchant and stored in all offline terminals. Each token signer is a process running on a physically-separated hardware. All signers are listening to a smart-contract event triggered by the \emph{token request function} of the private smart contract. This configuration allows the signers to communicate with the blockchain using a pull-only protocol, without listening on any ports.

\noindent\textbf{Offline Point of Sale}
The offline point of sale (OPOS) stays offline during the normal operation. Each OPOS stores its own private key and the full list of signers' public keys. Each OPOS has its own list of merchandise and pricing policy. If the list of signers' keys is modified in the system, e.g., a new signer is added to the system for extra redundancy, this information must be manually updated on each OPOS. No other change in the merchant infrastructure requires updating OPOS. If OPOS is compromised, i.e., its private key is stolen, the corresponding public key should be removed from or blacklisted in the private smart contract.

\noindent\textbf{Client}
\volgapay client is a mobile device, which can establish a client-OPOS channel with an OPOS in proximity. The client also has access to the merchant's private blockchain through a client-blockchain channel. For purchases, the clients do not access the public blockchain. 

\subsection{Communication Channels}
\volgapay has four classes of communication channels: blockchain grafting channel, client-OPOS channel, client-blockchain channel, and signer-blockchain channels, as shown in Figure~\ref{fig:design}.

\noindent\textbf{Blockchain Grafting Channel}
The blockchain grafting channel virtually links the address spaces of two independent blockchains. Specifically, it connects the public and private blockchains by storing the address of the private smart contract in the public smart contract.

\noindent\textbf{Client-OPOS Channel}
\volgapay client establishes a local bi-directional simplex channel with an OPOS in proximity. We assume the channel has limited capacity, possible low bandwidth, and is not persistent. For our prototype, as described in Section~\ref{section:implementation}, we use two-way QR-code scanning as client-OPOS channel. However, alternative channels can be established, including electromagnetic, Bluetooth, etc.

\noindent\textbf{Client-Blockchain Channel}
The client-blockchain channel allows the OPOS to get necessary updates on the state of the micropayment channel. While OPOS remains offline, the client plays the role of a proxy to the merchant infrastructure, through which the OPOS receives a receipt of successful payment in the form of a verifiable token. For the most common use cases of \volgapay, e.g., vending machine purchase, it is necessary for the client to have access to the merchant infrastructure while interacting with OPOS.

\noindent\textbf{Set of Signer-Blockchain Channels}
Each signer is an independent network node subscribed to the token request events of all the private smart contracts through a signer-blockchain channel. This arrangement allows the signers not to listen to any ports and remain anonymous, and therefore significantly limit the exposure to potential cybersecurity threats. In order to serve multiple clients simultaneously, the signer-blockchain channels must maintain sufficient available bandwidth, which is discussed in Section~\ref{subsection:scalability}. 

\subsection{\volgapay Protocols}\label{subsection:protocol}
\volgapay protocols define communication procedures between the participants of the system. \volgapay includes four major protocols: 1) \textit{contract initiation and deposit}, which establishes relationship between client and merchant; 2) \textit{transaction protocol}, which describes the procedure of making transactions; 3) \textit{payoff and refund}, which defines the conversion of the micropayment channel state into a fiat cryptocurrency; and 4) \emph{payoff verification}, which verifies the correctness of the payoff amount. None of the four protocols require the client to share any sensitive or personal information; thus, data privacy and identify anonymization are inherently guaranteed by our blockchain-based design.

\noindent \textbf{Contract initiation and deposit.}
This protocol includes the following sequence of steps:
\begin{enumerate}
    \item The client requests a contract from the merchant, e.g., through the merchant's website or using a mobile app, and sends to the merchant the heads of hash chains in the set $\Upsilon$;
    \item The merchant prepares a pair of smart contracts for the client---public and private, and sends the client the address of the public contract;
    \item The client verifies the code of the public smart contract, extracts the address of the private smart contract, and deposits funds on the public smart contract.
\end{enumerate}

\noindent \textbf{Transaction protocol.}
The purchase protocol includes the following steps: 

\begin{enumerate}
    \item A client approaches an OPOS and initiates a purchase, e.g., by selecting an item with description $D$ and price $\delta$ from the list of merchandise, as exemplified in Figure~\ref{fig:opos1};
    \item The OPOS generates a random unique transaction ID $\eta$, and delivers the tuple $(\eta, \delta, D, K(\eta, \delta))$ to the client;
    \item The client converts the item price $\delta$ into the $\tau$-adjusted price $\phi$, using the variable $\tau$ stored in the private smart contract;
    \item The client calls the \emph{token request function} of the private smart contract, providing $(R_{1}, R_{2},...,R_{\xi}), \eta, \delta$, and $K((\eta,\delta))$ as arguments;
    \item The merchant's private smart contract first verifies price, payment, balance, and the authenticity of the signature of the OPOS, and then stores the tuple of payment hashes in the smart contract. It updates the smart contract balance, and adds current payment amount to the sum of previous payments $\tau$, which is stored in the private smart contract. Finally, it triggers the event $E(\eta, \delta)$;
    \item Each token signer responds to the event $E(\eta, \delta)$ by storing the signature $\Gamma_{i}(\eta,\delta)$ in the private smart contract;
    \item The client waits until the private smart contract accumulates a minimum required number of signatures, and then it serializes these signatures into a tuple (which is the payment token), and delivers this tuple to the OPOS;
    \item The OPOS verifies whether the signatures are distinct, valid, and produced by legitimate token signers, and then delivers the purchased item to the client.
\end{enumerate}

\noindent \textbf{Payoff and refund.} 
The merchant first verifies the payment hashes in the private smart contract released by the client, and then calls a payoff function of the public smart contract with the inputs of payment hashes, payment amount, and optional client's payoff verification signature (as described below). The unused funds deposited in the smart contract can be delivered to the client as a refund. The implementation specifics of payment and refund are determined by the merchants' policies. 

\noindent \textbf{Payoff verification.}
\volgapay's polynomial multi-hash chain price representation creates a potential caveat for the merchant to learn more hashes from the hash chains than the desirable number of hashes determined by the value of $\phi$, as described in Section~\ref{subs:polymult}. \volgapay enforces the correctness of the payoff amount by adding a payoff verification signature, $sig(\phi)$, generated and signed by the client to authorize an exact transaction payment amount. The signature will be stored as an additional variable in the private smart contract and updated after each successful token request. The signature and corresponding $\phi$ will be verified by the public smart contract during the payoff process: if the merchant's requested payment deviates from the correct payment on the blockchain, the payoff will be subsequently rejected. We leave the implementation of this feature as future work.

%% file: implementation.tex
\section{System Implementation}\label{section:implementation}
We have implemented the \volgapay payment system. \volgapay prototype includes two independent OPOS emulators and two Android clients. The total size of implementation includes 6,100+ lines of code. 
The details of different parts of the prototype are discussed below.

\begin{figure}
\minipage{0.47\textwidth}
\centering
  \includegraphics[height=2.4in]{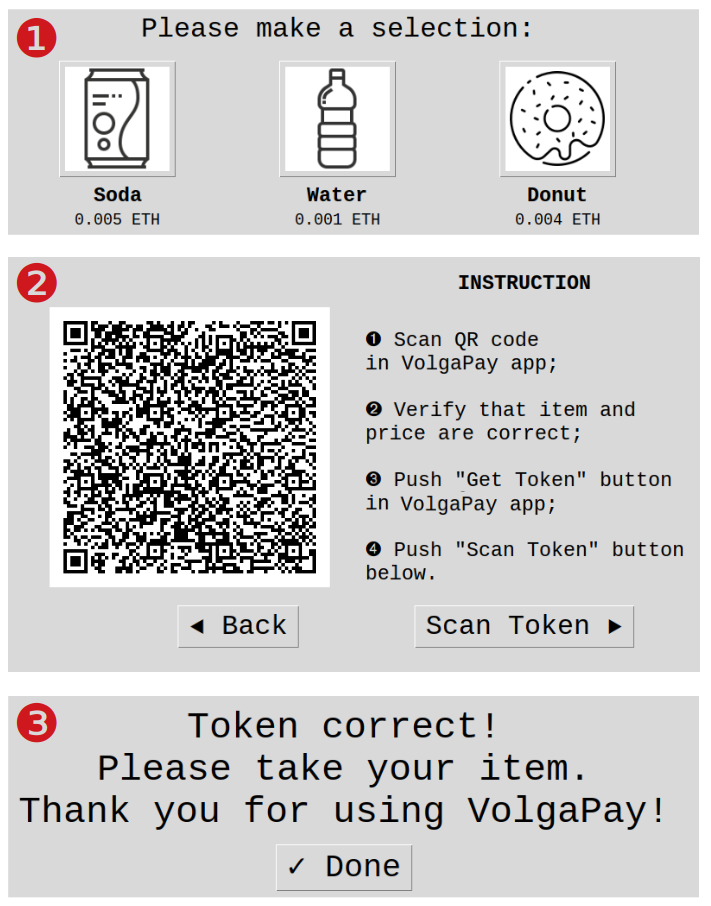}
  \caption{OPOS interface transition.}\label{fig:opos1}
\endminipage\hfill
\minipage{0.47\textwidth}%
\centering
  \includegraphics[height=2.4in]{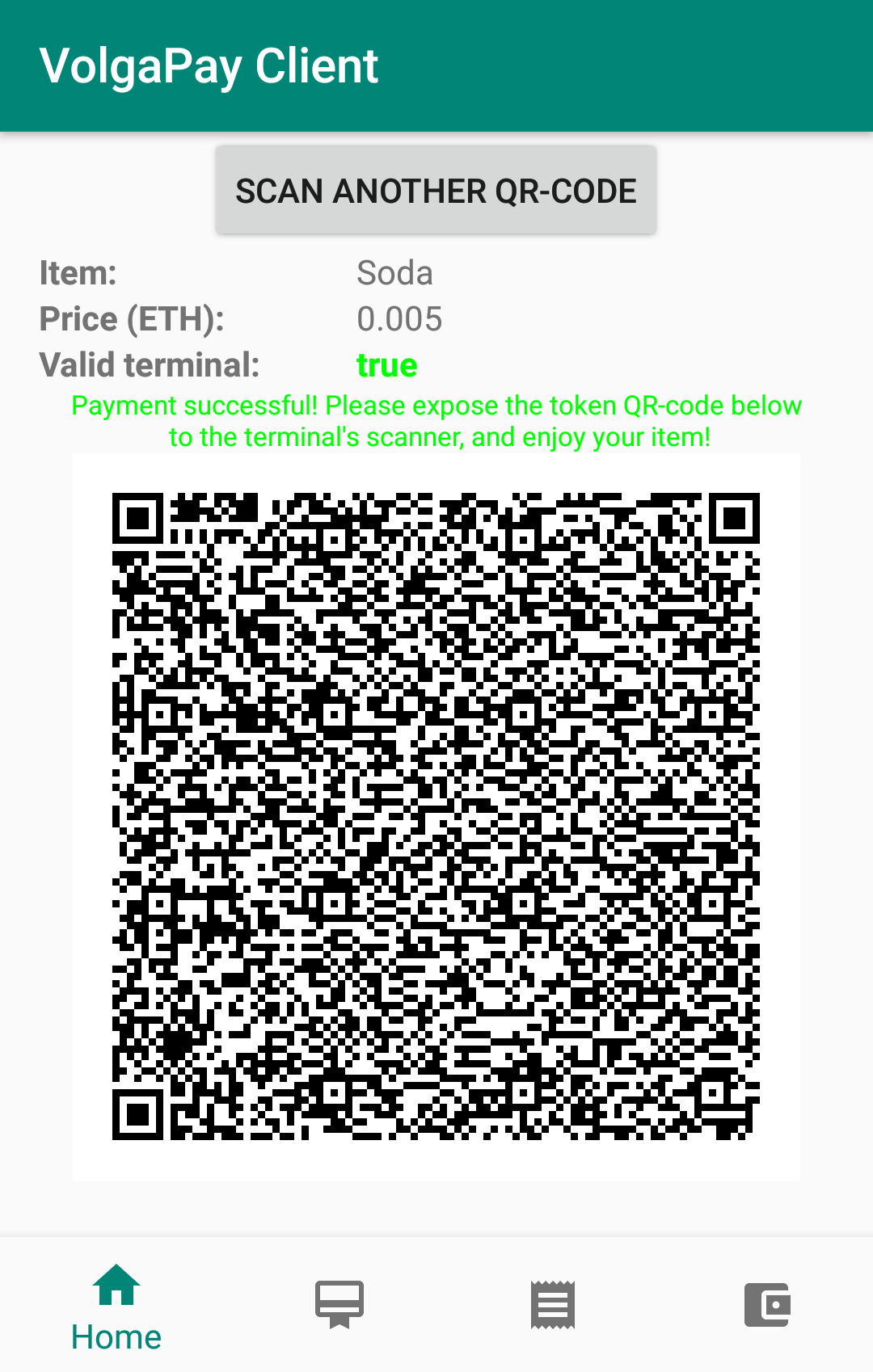}
  \caption{\volgapay Android client.}\label{fig:client1}
\endminipage
\end{figure}

\noindent \textbf{Offline Point of Sale (OPOS).}
We simulate two OPOS using two sets of Raspberry Pi 3B+ as computing modules, mini-touchscreen for user interaction, and Raspberry Pi camera for reading QR-codes. We use Python 3.6, TkInter, and Web3.py to implement the OPOS software. Both the terminals are offline. 

Figure~\ref{fig:opos1} shows a three-step user interface transition of a purchase. The first step is the selection of the item: when the user pushes the touchscreen button, the terminal generates a unique transaction ID, and produces a QR-code. The QR-code contains the transaction ID $\eta$, item price $\delta$ in \emph{wei}, item description $D$, and signature of the tuple $(\eta,\delta)$. During the second step, the client scans the QR-code with the Android app, requests a token, produces QR-code for the token, and pushes the ``Scan Token" touch button on the screen of OPOS, which activates the QR-code reader. Finally, when the QR-code is scanned, the OPOS verifies the signatures, and delivers the item to the client.

\noindent \textbf{Android client.}
We implement a \volgapay Android client using Android SDK and Web3j. Figure~\ref{fig:client1} shows the screen of the Android client right after obtaining a payment token from the private blockchain. Curious readers please refer to our demo video \url{https://youtu.be/rjIhDD2yi5I} for more information.

\noindent \textbf{Mainnet smart contract.}
The public smart contract, which we test on both Ropsten and Mainnet networks, uses preset hash chain heads for executing payoff to the merchant and refund to the client. For our prototype, we use the following parameters: base currency--Ether, $\Theta = 10$, $g = 10^{13}$, $\xi = 7$, which means that we use 7 10-hash-deep chains, with minimal price unit equal to 10 \emph{Szabo}, and maximum supported contract balance of 99.99999 \emph{Ether}.

\noindent \textbf{Kovan smart contract.}
We use Kovan Testnet for \volgapay prototype to simulate the private blockchain. \textbf{The code of the private smart contract, written in Solidity, is available through this link:} \url{https://github.com/seitlab/volgapay/blob/main/smartcontract.sol}. The \texttt{requestToken()} function performs the validation of client identity, the authenticity of the OPOS, the validity of the payment, and available balance. If all the parameters are verified, the private smart contract decreases the balance of the smart contract, updates the base price $\tau$, updates the payment hashes, and emits a signer-activation event.

\noindent \textbf{The signers.}
In our evaluation, we use a network of 10 signers. To demonstrate that the signers are not sensitive to latency and performance, we deploy them on low-tier basic servers (1 CPU/ 1 GB RAM / 25 GB HDD / Ubuntu 18.04.2 x64), located all across the world. Our prototype requires 5 valid signatures to form a payment token, i.e., $M=5$. The event listeners are written in JavaScript using Web3.js library. When the token request event is issued by the smart contracts, all the ten signers respond by signing the serialized tuple of transaction ID and item price, and write the signatures in the smart contract.

%% file: evaluation.tex
\section{Evaluation}\label{section:evaluation}
Using the prototype, we perform evaluation of \volgapay based on three criteria---delays, communication overhead, and cost of fees. Our prototype is not optimized for production use. Further improvement is possible through software optimization, choice of hardware, data compression techniques, network configuration, and selection of cryptographic routines. In the evaluation, each OPOS is implemented on a separate Raspberry Pi Model 3B+~\cite{rpi3bplus}; each signer has the following configuration: 1 CPU, 1 GB RAM, 25 GB HDD, Ubuntu 18.04.2. As clients, we use Huawei Mate 9 and Motorola Moto G5 Plus, both with Android Nougat. 

\subsection{Delays}
Here, we measure two types of delays, which determine the feasibility of \volgapay: token request time and payment request delay on public blockchain.

\subsubsection{Transaction Latency of Traditional Payment Methods}
We conduct a field experiment to measure a payment transaction latency of different types of traditional self-service terminals using Chase VISA debit card and CapitalOne MasterCard credit card, using both magnetic stripe and RFID technologies. We made 2 purchases at 2 different self-service gas stations, 4 purchases at 3 vending machines, 2 purchases at two parking meters, 2 balance checks, 2 cash withdrawals, and 1 PIN verification at 2 ATMs belonging to 2 different banks. We measure the transaction delay between applying the payment and receiving a payment acknowledgement using a handheld stopwatch. In order to adjust for human response delay, we subtract 1 second from each reading rounded to the closest second. The results, shown in Table~\ref{table:tokendelays}, exhibit a significant variance, with the transaction times ranging from 5 to 21 seconds, with standard deviation of 5.3, and mean average of 11.7 seconds.

\subsubsection{Token Request Time}
We evaluate the token request time by measuring the time delay in milliseconds between pushing the ``Get Token" button in the client Android App and the appearance of the token QR-code on the screen of the client. We measure the delay under three different types of Internet connection of the client: fast WiFi, LTE, and 3G. For each of the three connection types, we perform 20 probes, and record the mean average, standard deviation, and the speed profile of each connection\footnote{For speed profile, we measure the speeds using Ookla Speed Test and M-Lab Speed Test, and get their average.} in Table~\ref{table:tokendelays}. The experiment shows the average token request time, ranges between 9 and 12 seconds.

\begin{table}
\begin{center}
\caption{Token request delays with different client connections. The transaction latency of \volgapay is comparable to the credit card transaction delays with traditional payments (ranging from 5 to 21 seconds in our field experiment).}
\label{table:tokendelays}

\begin{tabular}{|c|c|c|c|c|c|}
\hline
\textbf{Client}&\multicolumn{3}{|c|}{\textbf{Avg. link bandwidth}}&\textbf{Delay}&$\mathbf{\sigma}$ \\
\cline{2-4}
\textbf{connection}&Down&Up&Ping&\textbf{(ms)}& \\
\hline

WiFi&39.67&18.02&10.5&9,399&1,837 \\
\hline

LTE&13.57&12.99&26&10,026&1,769 \\
\hline

3G&2.54&0.90&80.5&11,618&1,716 \\
\hline
Traditional self-service with credit cards & --- & --- & --- & 11,700 & 5,300 \\
\hline

\end{tabular}
\end{center}
\end{table}

\subsubsection{Public Smart Contract Calls}

We measure the delays of public smart contract calls on Ropsten Testnet and Etheretum Mainnet networks. Each measurement is repeated 10 times, and the average results recorded in Table~\ref{table:publicchaindelay}. The results show that the Ropsten network, despite its similarity with the Mainnet, does not always produce similar execution delays. The measurements also show that all the smart contract calls complete within 30 seconds on average. The results prove that the \volgapay transactions introduce reasonable delays.

\begin{table}
\begin{center}
\caption{Average public smart contract delays on Ropsten and Mainnet networks using Infura API, based on 10 measurements, with gas price 2.2 Gwei.}
\label{table:publicchaindelay}

\begin{tabular}{|c|c|c|c|c|}
\hline
\textbf{PHCS}&\multicolumn{2}{|c|}{\textbf{Ropsten Testnet}}&\multicolumn{2}{|c|}{\textbf{Ethereum Mainnet}} \\
\cline{2-5} 
\textbf{Operation} & \textbf{\textit{delay (ms)}}& \textbf{\textit{$\sigma$}}& \textbf{\textit{delay (ms)}}&\textbf{\textit{$\sigma$}} \\
\hline

Bytecode check&756&371&10,920&698 \\
\hline

Balance check&79&28&700&12 \\
\hline

Deposit&28,168&13,874&22,555&20,823 \\
\hline

Set 7 hash chain heads & 26,467 & 11,547 & 26,491 & 25,222 \\
\hline

Retrieve hash chain heads & 489 & 41 & 554 & 65 \\
\hline

Retrieve stored public key & 175 & 46 & 236 & 205 \\
\hline

Payoff&23,333&28,338&21,413&16,019 \\
\hline

Refund&20,031&15,400&27,898&25,529 \\
\hline

Contract deployment & 24,481 & 10,913 & 20,231 & 9,623 \\
\hline
\end{tabular}
\end{center}
\end{table}

\subsection{Communication Overhead}
\subsubsection{Token Request}
\begin{figure}
\captionsetup[subfigure]{justification=centering}
\centering
\begin{subfigure}{.47\linewidth}
  \centering
  \includegraphics[width=2.4in]{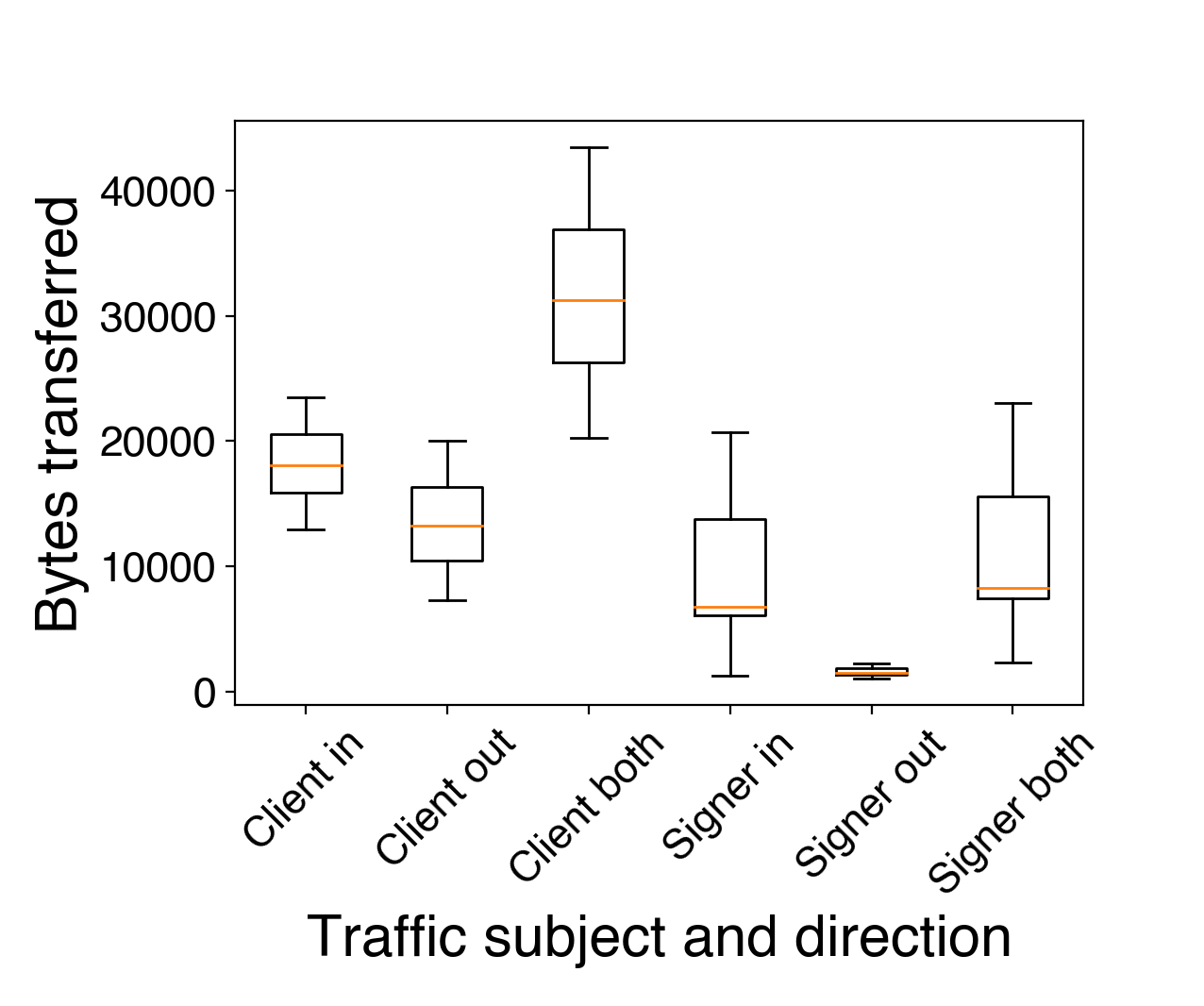}
  \caption{Avg. token request communication overhead over 30 measurements.}
  \label{fig:token-overhead}
\end{subfigure}~~~
\begin{subfigure}{.47\linewidth}
  \centering
  \includegraphics[width=2.6in]{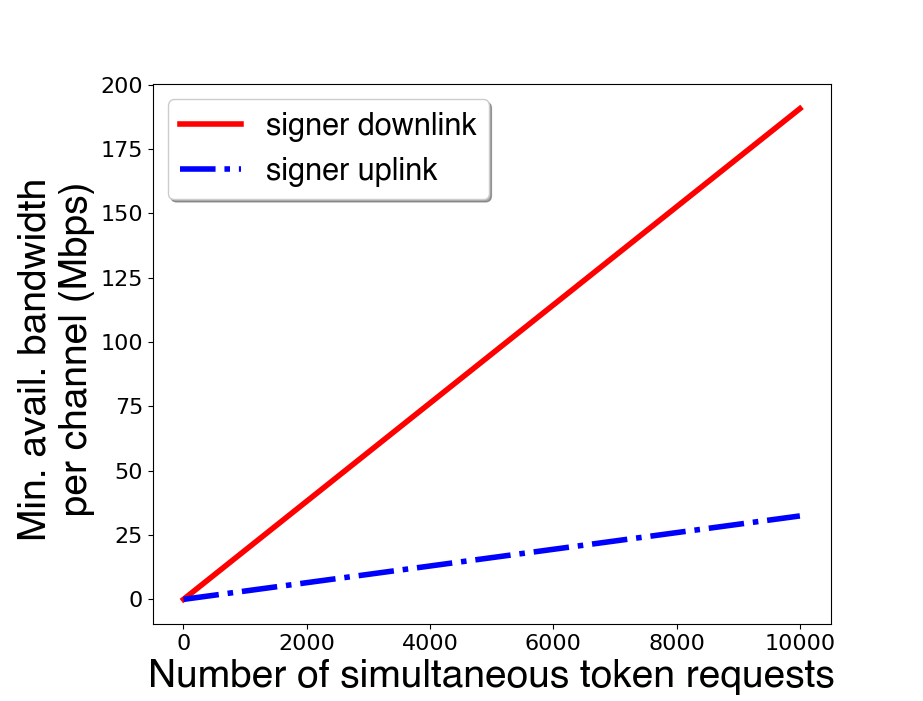}
  \caption{Min. avail. bandwidth requirement per signer-blockchain channel.}
  \label{fig:scal1}
\end{subfigure}%

\caption{VolgaPay communication overhead and associated throughput.}
\label{fig:throughput}
\end{figure}

In our prototype, we use Infura API for communication with Kovan network. The summary of the communication overhead through the client-blockchain channel over 30 measurements is summarized in the left half of Figure~\ref{fig:token-overhead}. The result shows that the overall communication overhead does not exceed 50 Kbytes, and the incoming traffic volume is slightly larger than the outgoing one. The right half of the graph delivers the summary of communication overhead over signer-blockchain channels, which shows the total overhead per signature rarely exceeds 20 Kbytes, and the incoming traffic volume is also much larger than the outgoing one. 

\subsubsection{QR-Codes}
The ASCII-based QR-code produced by the OPOS in our prototype includes the following fields: transaction ID, price in \emph{wei}, description, and signature. The total length of the sequence encoded in the OPOS QR-code ranges between 175 and 222 bytes. The length of the second QR-code (i.e., payment token) is determined by the parameter $M$ (i.e., the number of required signatures) of the system. For \volgapay prototype, we use $M=5$ with \texttt{0x}-prefixes and 1-byte delimiters (for error checking), allotting to 664 bytes per token.

\subsection{Fees}
Table~\ref{table:gasfees} shows the gas fees of \volgapay prototype's public smart contract implementation over 10 measurements on Ropsten and 10 measurements on Mainnet network. As shown in the table, the operations are cheap, and the most gas-consuming one is contract deployment. Note that smart contract deployment is a one-time cost, which is less than \$1. 

\begin{table}
\begin{center}
\caption{Gas fees, based on 20 measurements.
}
\label{table:gasfees}

\begin{tabular}{|c|c|c|c|}
\hline
\textbf{Smart contract}&\textbf{Avgerage}&$\sigma$&\textbf{Approximate} \\
\textbf{operation}&\textbf{gas used}&&\textbf{USD cost}$^{\mathrm{a}}$ \\
\hline

Set addresses&21,040&0&\$0.012 \\
\hline

Set heads&72,263&65&\$0.040 \\
\hline

Payoff&55,212&53&\$0.030 \\
\hline

Refund&30,001&0&\$0.017 \\
\hline

Contract deployment&922,915&60,545&\$0.508 \\
\hline

\multicolumn{4}{l}{$^{\mathrm{a}}$With the gas price 2.2 Gwei.}
\end{tabular}
\end{center}
\end{table}

\subsection{Scalability}\label{subsection:scalability}
\noindent \textbf{Number of simultaneous token requests.}
Let us assume that the available bandwidth of all signer-blockchain links is approximately the same, and the maximum time allotted for one signature response equals the average time $T_{b}$ (in seconds) of sealing one block in the private blockchain. Then, the minimum available bandwidth $B_{min}$ for one signer (in bps) is $B_{min} = {{N_{R} \cdot {L}} \over {T_{b}}}$, where $L$ is communication overhead per request in bits, and $N_{R}$ is the number of simultaneous token requests. Using this formula and the communication overhead measurements gathered from the \volgapay prototype (see Figure~\ref{fig:token-overhead}), we can demonstrate (see Figure~\ref{fig:scal1}) that a channel with 100 Mbps (200 Mbps) bandwidth is able to support more than 5,000 (10,000) simultaneous token requests.

\noindent \textbf{Number of clients.}
The number of clients in \volgapay is not limited by any practical parameter of the system. However, it is important to note that as the number of clients increases, the probability of exceeding the bandwidth capacity of the signer-blockchain channel grows.

\noindent \textbf{Number of OPOS terminals.}
The number of OPOS terminals affects two parameters in the private smart contract: the set of OPOS public keys, and the amount of computation required to verify the identity of a terminal. We modify the smart contract by adding $10^3$ random OPOS keys, which increases the number of OPOS identity checks from 2 to 1,000 in the token request routine. Then, we re-evaluate the token request delay by running 20 more token requests simulating the worst-case scenario. The results indicate similar delays, which do not exceed 10 seconds on average. Therefore, for each merchant, \volgapay can support at least 1,000 OPOS terminals without degrading performance.

%% file: security.tex
\section{Security Analysis}

Here, we account for the possible reasonable scenarios in which important components (participants or channels) are compromised by an adversary. If several types of attacks exist for achieving the same malicious outcome, we elaborate on the outcome, e.g. ``gaining full access".

\noindent \textbf{Malicious merchant.}
Malicious merchants try to steal funds from the clients. Note that the clients' funds in public smart contract are guarded by the public blockchain and their respective private keys. The only feasible malicious activity would be to receive payment without providing a valid token, in which case the client will lose payment for a single purchase. Given the cost of infrastructure setup, the merchants are unlikely to pursue such attacks.

\noindent \textbf{Full access to OPOS.}
Each OPOS stores its own private key, which is used for confirming its belonging to a certain merchant. If the key is stolen, it can be temporarily used for the deployment of a fake OPOS, but it does not allow the attacker to steal any funds since OPOS does not process payments. Therefore, the incentives to steal the private key of an OPOS is limited only to retaliation and vandalism.  In order to protect the private key from theft, it is recommended to store the key only in the address space of RAM of the legitimate \volgapay process. Additionally, if an attacker gains full physical access to an OPOS, all the merchandise of the terminal can be stolen bypassing the token verification procedure. This, however, can be prevented by enforcing physical security, which goes beyond the scope of this work.

\noindent \textbf{Full access to the client app.}\label{fullaccessclient}
The client app can be designed not to store the private key used in public smart contract, or to store it securely encrypted, with mandatory passphrase solicitation before each use. As a result, we exclude the possibility of stealing the private key used for public smart contract even when the adversary gains access to the client software. Thus, the only valuable asset that can be stolen from the client is the set of hash chain heads, and the authentication private key for the private smart contract: the combination thereof can be used to make purchases on behalf of the client, but does not allow to steal funds because fund management is performed via the private key of the public smart contract.

\noindent \textbf{Full access to one of the private chain sealers.} 
If one of the sealers of the private blockchain is compromised, there is no immediate threat to the system. Moreover, depending on the specifics of the private blockchain's consensus protocol and the size of the clique or quorum, the blockchain remains intact even if several sealers are compromised. In most cases, it requires to compromise 50\% of the sealers to attack the private blockchain. However, a compromised sealer likely means one of the private keys used for the consensus is stolen. Although there is no immediate danger, the sealer must eventually be replaced, which will require creating a new genesis block, restarting blockchain, re-deployment of private smart contracts, and changing the grafting links in corresponding public smart contracts, which can be done automatically during a scheduled maintenance.

One way to mitigate the threat associated with compromised private blockchain sealers is to use reputable outsourced PoA blockchain services, or to share blockchain between several mutually trusted merchants. Another solution is to add additional IP authentication to the boot nodes, which grants access to the blockchain's P2P network only to peers from certain IP addresses; this solution, however, requires to run a sufficient number of boot nodes in order to prevent an eclipse attack~\cite{wust2016ethereum}.

\noindent \textbf{Denial of service attacks.}
Although denial-of-service attacks do not yield any immediate fund gain for an attacker, they can be used as a means of retaliation or vandalism. Private blockchain sealers and \volgapay signers are possible targets to this type of attack. The design of \volgapay allows to keep the IP addresses of both signers and sealers in secret; and if this information still gets revealed, to seamlessly replace the IP addresses without changing blockchain-level identities (keys). This is possible because sealers in blockchain are identified by their keys, not by their IP addresses.

\noindent \textbf{Local channel sniffing and spoofing.}
Among many ways of maintaining the client-OPOS channel, the QR-code exchange is the most feasible and popular approach. The visual channel is subject to eavesdropping attacks, and its payload limitations make it infeasible to secure, e.g., through a Diffie-Hellmann key exchange. Thus, we can assume that the information in the channel can be read (sniffing) and substituted (spoofing). Because of the physical presence of the client at the OPOS for the entire duration of transaction, the sniffing or spoofing of any of the two QR-codes does not yield any gain to the attacker other than creating a minor annoyance to the client.

\noindent \textbf{Man-in-the-middle-attacks.}
The resistance to man-in-the-middle (MITM) attacks in all the \volgapay channels, except the client-OPOS channel described above, can be achieved by establishing encrypted communications. Since most of the channels, such as secure socket or RPC-calls to blockchain, are secure by default, the possibility of such an attack is very low. However, even if an MITM attack is successfully conducted, the potential harm can only be experienced by the client, whose payment token could be blocked by an attacker. There is little incentive for an attacker to do so, as the token validity is limited by the current transaction session.

%% file: conclusion.tex
\section{Conclusion}
In this paper, we presented a new blockchain-based tokenized payment system, \volgapay, to address the numerous practical challenges towards the deployment and use of offline payment terminals, such as security, privacy, trust in merchant, and significant infrastructure cost. \volgapay incorporates multi-hash chain-based micropayment channels and  blockchain grafting to strike a balance in security and efficiency, by leveraging the security and trustworthiness of public blockchain, as well as the speed and cost-efficiency of private blockchain. We implemented the \volgapay payment system and evaluated all the parameters affecting its practical feasibility. Our evaluation shows that the system is fast, cost-efficient, and scalable. Most importantly, through a comprehensive security analysis, we demonstrated that \volgapay is resistant to a variety of cyber-attacks: if any component of the system is compromised, the scope of harm will be minimized, and the threat will not propagate to other components.

%% file: acknowledgement.tex
\section*{Acknowledgement}\label{sec:acknowledgement}
We would like to thank the anonymous reviewers for providing valuable feedback on our
work. 
This work was supported in part by National Science Foundation grants CNS1950171 and CNS1949753.